# Cavity Ferromagnetic Resonance Study of Acoustic and Optic excitations in Ru/Cr/Co and Ru/Co multilayers.

P. Ntetsika[1], G. Mitrikas[2], G. Litsardakis[3], I.Panagiotopoulos[1,4]

1   Department of Materials Science and Engineering, University of Ioannina, Ioannina 45110, Greece
2   Institute of Nanoscience and Nanotechnology, National Centre for Scientific Research-Demokritos, Athens, Greece
3   Laboratory of Materials for Electrotechnics, Department of Electrical and Computer Engineering, Aristotle University of Thessaloniki, Thessaloniki, Greece.
4   Institute of Materials Science and Computing, University Research Center of Ioannina (URCI), 45110 Ioannina, Greece

**Abstract**

Two series of $[Ru_{10}/Co_x)]_{12}$ and $[Ru_6/Cr_3/Co_x)]_{12}$ x=16-60 easy-plane anisotropy multilayers (all thicknesses in Å), prepared by sputter deposition, are studied by cavity FMR. The acoustic modes are excited by setting the RF field perpendicular to the saturating in-plane field. Their resonance fields are in good agreement with the values predicted for acoustic modes by macrospin models, using the interlayer RKKY exchange and anisotropy fields derived by the magnetic measurements. The resonance fields of the modes excited by setting the RF field parallel to the saturating field, are lower than those expected for optical modes. This and could attributed to the existence of hybridized mixed modes, according to results of micromagnetic simulations which also show that the mode mixing is related to the inhomogeneous magnetization profile along the multilayer thickness.

KEYWORDS: Synthetic Antiferromagnet, Magnetization Dynamics, FMR, Magnon Coupling, Magnetic multilayer, Sputter deposition.

## 1. Introduction

Synthetic antifferomagnets (SAFs) are based on the oscillatory interlayer exchange coupling of thin magnetic layers through metal which is typically mediated through the conduction electrons with the Ruderman–Kittel–Kasuya–Yosida (RKKY) mechanism [1]. The low field tunability of the magnetic state in SAFs opened the route for the discovery of the giant magnetoresistance phenomena and their implementation in many spin-valve and magnetic tunnel junction designs [2]. It is the same tunability that makes them attractive for other applications such as tunable magnonics [3,4,5] and terahertz nano-oscillators [6,7,8]. In early studies, the focus was on the optimization of the SAFs for spin valves, the study of dynamic properties was used mainly as a technique to derive anisotropies and interfacial coupling strength [9]. Today there is a renewed interest in SAF systems for their dynamic properties themselves and due to their possible incorporation in synthetic antiferromagnetic spintronic structures [10].

The antiferromagnetic interlayer exchange coupling in SAFs gives rise to two distinct modes (acoustic / optical) distinguished according to the correlation of the precession of the magnetic moments between adjacent magnetic layers. In Fe/Cr multilayers besides an acoustic branch, several additional modes were observed under parallel excitation of resonance which correspond to excitation of standing spin waves with wave vectors perpendicular to the film plane [11,12]. As their names indicate, for the acoustic modes the resonances approach zero at zero field, while the optical modes possess useful high frequencies at zero-applied field [13]. The frequencies of these modes, at a specific applied field, depend on the magnetic state and therefore on uniaxial magnetic anisotropy ($H_K$) and antiferromagnetic interlayer exchange coupling ($H_{ex}$), which provide control parameters of the dynamic properties. Several works are focused on these aspects of SAF systems as FeCoB/Cr/FeCoB [14], FeCoB/Al$_2$O$_3$/FeCoB [15], Pt/Co/Ir/Co/Pt [16], CoFeB/Ru/CoFeB [17,18]. A study based on all-optical pump-probe technique showed that the dynamic exchange coupling increased only the damping of the optical mode owing to the spin-pumping effect at the CoFeB/Ru interfaces [19]. It has been shown that a higher optical mode frequency is obtained at the second antiferromagnetic coupling peak [13]. Inversely, the frequencies of the modes can be used to extract the coupling and anisotropy parameters and has been used both in early [20] and recent studies [6]. As the frequencies of the optical and acoustic modes depend on the applied field, there is a point where they tend to coincide. The optical and acoustic spin-wave modes get hybridized at these degeneracy points [4,5,21]. The mode coupling is reported to be mediated by the dipolar fields generated by the magnetization motion of spin waves and the out-of-plane tilt angle [4]. Therefore it increases with the wave number of exited spin waves and the angle between the external magnetic field and spin-wave propagation directions [5] and can be enhanced applying an out of-plane bias field [22] or constructing a structurally asymmetrical SAF [21].

Here we present a Ferromagnetic Resonance (FMR) study of two series of [Ru$_{10}$/Co$_x$)]$_{12}$ and [Ru$_6$/Cr$_3$/Co$_x$)]$_{12}$ x=16-60 multilayers (all indices inside the brackets denote thickness in Å, the index 12 refers to the repetitions of the period of the stack) prepared by sputter deposition having easy-plane anisotropy and zero anisotropy within the plane. Co/Cr heterostructures have received much less attention than the Fe/Cr since there is a crystal structure mismatch between Cr and Co [23]. The use of chromium gives an extra degree of freedom in tailoring exchange coupling and anisotropy independently as both have interfacial contributions that scale inversely with the layer thickness. Furthermore, the use of chromium, along with ruthenium, in the RKKY layer introduces an asymmetry to the deposited multilayer along the stacking direction.

## 2. Experimental details

The multilayered [Ru$_{10}$/Co$_x$)]$_{12}$ and [Ru$_6$/Cr$_3$/Co$_x$)]$_{12}$ (with x=14-60, all thicknesses in Å) films have been deposited on rotating substrates, at room temperature by magnetron sputtering, using a multi-source deposition system. The values of the Ru and Ru/Cr layer thickness were chosen to maximize the RKKY exchange interactions in each case. The Co, Cr (7.62 cm diameter) and Ru (5.08 cm diameter) sources are in confocal geometry: i.e., pointing at an angle of 45° to the (horizontal) substrate plane, which is rotated during the deposition. The target to substrate distance is 15 cm. Prior to the deposition, the chamber was evacuated to a base pressure better than 5×10$^{-7}$ Torr and the process gas (Ar) pressure during deposition was 3.5 mTorr. Co has been deposited at a rate of 0.75 Å/s by applying a DC power of 100 W, chromium at 1.15Å/s with 130 W DC, and Ru at 0.40 Å/s using 120 W RF. The samples were sputtered on thermally oxidized Si(100) wafers. The magnetic measurements were done using a vibrating sample magnetometer (VSM) of Lakeshore Cryotronics Inc. (model 7312) with eight sensing coils (four per component) and a 20 kOe electromagnet and a QD Versalab VSM with 30 kOe field. The ferromagnetic resonance (FMR) measurements were performed at room temperature using a Bruker ESP 380E spectrometer equipped with a rectangular ER 4102ST or a dual-mode ER 4116DM cavity. The dc-field was always in the film plane. When the exciting RF field is parallel to the dc-field the excitation of optical modes for which the motion of the spins is symmetric with respect to the dc-field direction are favored. On the other hand, setting the RF field perpendicular to the dc favors the excitation of acoustic modes. The microwave frequency was measured with a HP 5350B microwave frequency counter. Spectra were obtained using 20.9 mW of microwave power, 100 kHz of modulation frequency, 1 mT of modulation amplitude, a field-sweep range of 450 mT and acquisition time of 167 s. The cavity frequencies were 9.76GHz for the RF(⊥) and 9.34 for the RF(//) measurements.

## 3. Magnetic Properties and FMR

Since the films are deposited on rotating substrates there is no anisotropy within the film plane. The shape anisotropy is the dominant term: for all the films the saturation field within the plane $H_\parallel$ is lower than the saturation field along the film normal $H_\perp$. In short, the anisotropy is easy plane but without any preferential orientation within the film plane. Due to the lack of anisotropy within the plane, the scissor-like magnetization state between successive Co layers (due to the competition between the interlayer AF coupling and dc magnetic field) is attained gradually without any spin-flop transition [24]. Therefore, the RKKY exchange field $H_{ex}$ can be estimated by the in-plane saturation field $H_\parallel = \frac{4J_{AF}}{M_s t_{Co}} = 2H_{ex}$. The factor 2 enters since each Co layer is RKKY coupled to two neighbouring layers and holds strictly only for an infinite layer stack. However, for our case with N=12 Co layers, the micromagnetic simulations (section 4) show that the effect of the two outer layers (which are coupled to only one layer) is negligible. For the measurements along the film normal except the shape anisotropy, magnetocrystalline ($K_{mc}$) and interfacial anisotropy ($K_S$) contributions must also be considered. Therefore, the total uniaxial anisotropy would be $K_{eff} = -\frac{1}{2}\mu_0 M_s^2 + K_{mc} + \frac{2K_S}{t_{Co}}$ and the field $H_K$ required for saturation against the anisotropy is $\mu_0 H_K = \mu_0 M_s - \frac{2K_{mc}}{M_s} - \frac{4K_S}{M_s t_{Co}} < \mu_0 M_s$. Thus, the values of $H_{ex}$ and $H_K$ are estimated by the saturation field parallel and perpendicular to the film plane $H_\parallel, H_\perp$ respectively using $H_\parallel = 2H_{ex}, H_\perp = H_K + 2H_{ex}$. Typical curves for [Ru$_{10}$/Co$_{30}$)]$_{12}$ and [Ru$_6$/Cr$_3$/Co$_{30}$)]$_{12}$ samples are shown in Fig.1. The obtained values are summarized in Table I along with the FMR resonance fields. The dependence of $H_{ex}$ and $H_K$ on the Co layer thickness is presented in Fig.2. The thickness dependences can be explained by the decreasing interfacial contributions as the thickness increase: the exchange field is mainly of interfacial origin so it is expected to decrease with thickness. Some deviations may arise from magnetostatic coupling. On the other hand, the main contribution to $H_K$, is the shape anisotropy which is reduced by the opposing interfacial contributions. Therefore, $H_K$ increases with Co thickness.

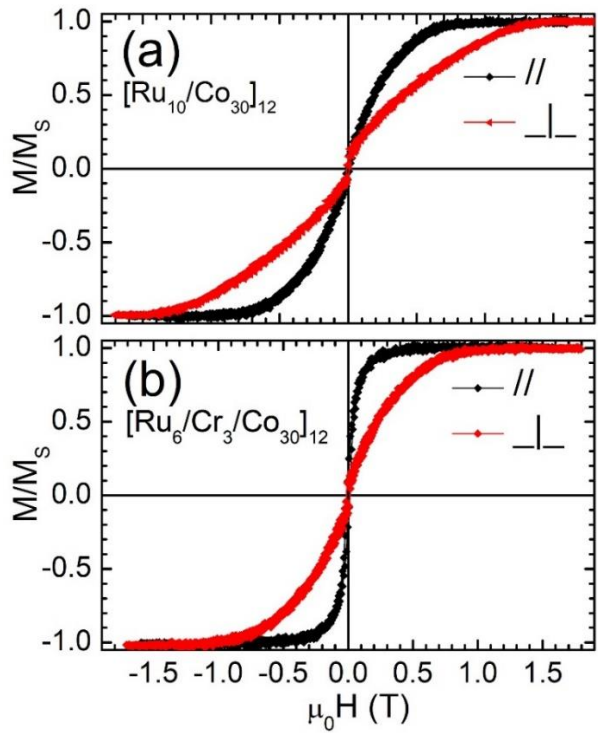

Fig.1 *Typical VSM magnetization curves for (a) [Ru$_{10}$/Co$_{30}$)]$_{12}$ and (b) [Ru$_6$/Cr$_3$/Co$_{30}$)]$_{12}$ samples. These curves are full loops from positive saturation to negative and back to positive. The paths of the two branches coincide*

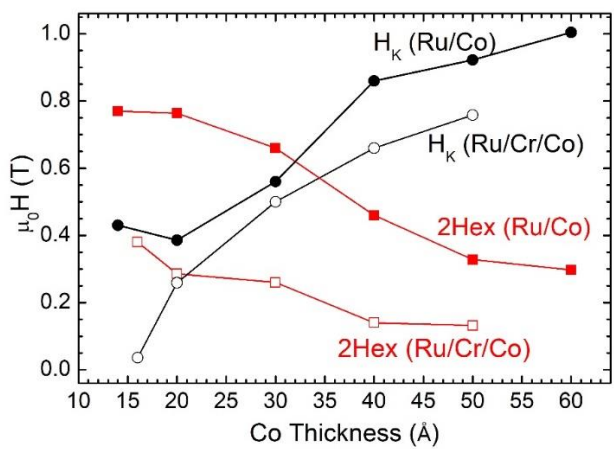

Fig.2 *Dependence of interlayer exchange field H$_K$ (squares) and anisotropy field H$_K$ (circles) on Co layer thickness for [Ru$_{10}$/Co$_{30}$)]$_{12}$ (solid symbols) and (b) [Ru$_6$/Cr$_3$/Co$_{30}$)]$_{12}$ series (open symbols).*

A 30 nm single layer Co film prepared under the same conditions shows a resonance at 0.126 T (Fig3(a)). The saturation field perpendicular to the substrate is $H_\perp = 0.96\ T$ whereas the magnetization value (considering only shape anisotropy) would imply $H_\perp = \mu_0 M_s = 1.46\ T$. Since for this sample there are not interfacial contributions, we assume a contribution from the magneto-crystalline anisotropy of $H_{mc} = 0.5\ T$, giving $K_{mc} = 290$ kJ/m$^3$ compared to the bulk value of 520kJ/m$^3$. Using the Kittel $f = \gamma \sqrt{H(H + H_{K,eff})}$ where $H_{K,eff} = H_\perp = 1.46\ T$ the resonance at 0.126 T corresponds to a value $\gamma = 26.5$ GHz/Tesla which is close to the value given for hexagonal cobalt [25].

For SAF multilayers, and fields lower than the $2H_{ex}$, the sample is unsaturated and the acoustic and optical resonances are expected at [4]

$$\omega_a = \gamma H \sqrt{1 + \frac{H_K}{2H_{ex}}} \quad , \omega_o = \gamma \sqrt{2H_{ex}H_K}\sqrt{1 - \left(\frac{H}{2H_{ex}}\right)^2} \ \text{(eqs.1)}$$

where $H_K$ is the effective anisotropy field which includes all magnetocrystalline/interfacial and shape anisotropy contributions. These equations imply that there is a field value $H^* = \sqrt{2}H_{ex}/\sqrt{1 + H_{ex}/H_K}$ for which the frequencies of the optical and acoustic branches coincide. For fields $H_\parallel > 2H_{ex}$ the sample is saturated, and the acoustic resonance coincides with that of a single ferromagnetic layer

$$\omega_a = \gamma \sqrt{H(H + H_K)} \quad \text{(eq.2)}$$

whereas the optical resonance is suppressed.

We present typical FMR spectra for the [Ru(6)/Cr(3)/Co(40)]$_{12}$ (Fig.3(b)) which has a typical behavior, as well as for the [Ru(10)/Co(40)]$_{12}$ sample (Fig.3 (c,d)) which has the largest number of resonances.

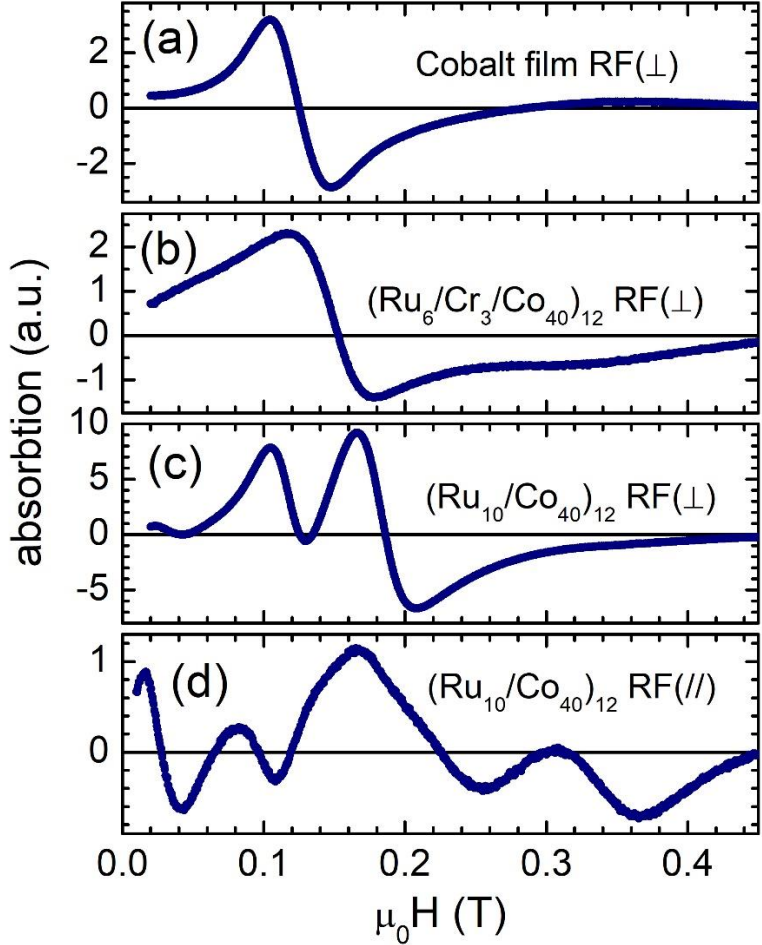

*Fig.3 FMR spectra for different mutlilayers measured with the RF perpendicular/parallel to in-plane dc field.*

Using the $H_\parallel, H_\perp$ obtained by the VSM measurements we derive the parameters $H_{ex}$ and $H_K$. Then using eqs.1 (with the $\omega$ set to the cavity frequency), the expected field values of the acoustic and optical FMR resonances ($H_{ac}$, $H_{opt}$) can be calculated. The application of an RF field perpendicular/parallel to the in-plane dc field (indicated as RF($\perp$)/RF(//) hereinafter) favours the excitation of acoustic/optical modes respectively [4,11]. Thus the $H_{ac}$ should be compared to the RF($\perp$) values and $H_{opt}$ to the RF(//) values. For the acoustic resonances there is a fair agreement between the calculated values and those observed by FMR for $\gamma = 32$ GHz/T, a value which is close to that obtained for Pt/Co/W multilayers prepared under the same conditions [26]. The

[Ru(10)/Co(14)]$_{12}$ sample did not give measurable signal in the FMR. The FMR measurements with RF(//) did not give resonances for any of the [Ru(6)/Cr(3)/Co(x)]$_{12}$ samples. For most of these samples $\gamma\sqrt{2H_{ex}H_K}$ is below of the RF frequency 9.77GHz of the cavity and therefore according to the second of the (eqs.1) cannot be observed.

The [Ru(10)/Co(20)]$_{12}$ samples gave resonances with RF(//) but at frequencies which are lower than the expected pure optical modes. Furthermore, despite having the RF along the magnetization direction, resonances at the expected acoustic modes are also observed. For [Ru(10)/Co(40)]$_{12}$ the higher resonance is at a value 0.34 T, between the calculated $H_{ac}$, $H_{opt}$ whereas the second one is very close to the $H_{ac}$. Similarly, the [Ru(10)/Co(30)]$_{12}$ sample resonance field value 0.39 T is between $H_{ac}$ and $H_{opt}$. For RF(//) resonances even below $H_{ac}$ appear. These findings can be explained by hybridization of optical and acoustic modes, according to the discussion of the results of micromagnetic simulations in the next section.

Table I
Saturation and Resonance Fields. All values are in Tesla.

| Layering | VSM | | macrospin model | | FMR | |
|---|---|---|---|---|---|---|
| | $\mu_0 H_{//}$ | $\mu_0 H_\perp$ | $\mu_0 H_{ac}$ | $\mu_0 H_{opt}$ | RF($\perp$) | RF(//) |
| [Ru(10)/Co(14)]$_{12}$ | 0.77 | 1.20 | 0.24 | 0.66 | - | - |
| [Ru(10)/Co(20)]$_{12}$ | 0.76 | 1.15 | 0.25 | 0.64 | 0.25 | - |
| [Ru(10)/Co(30)]$_{12}$ | 0.66 | 1.22 | 0.22 | 0.58 | 0.22 | 0.39 |
| [Ru(10)/Co(40)]$_{12}$ | 0.46 | 1.32 | 0.18 | 0.41 | 0.19, 0.12 | 0.34, 0.21, 0.09, 0.03 |
| [Ru(10)/Co(50)]$_{12}$ | 0.33 | 1.25 | 0.16 | 0.28 | 0.14 | 0.12 |
| [Ru(10)/Co(60)]$_{12}$ | 0.30 | 1.30 | 0.15 | 0.25 | 0.12 | 0.07 |
| [Ru(6)/Cr(3)/Co(16)]$_{12}$ | 0.38 | 0.42 | 0.24 | - | 0.25 | - |
| [Ru(6)/Cr(3)/Co(20)]$_{12}$ | 0.29 | 0.55 | 0.29 | - | 0.21 | - |
| [Ru(6)/Cr(3)/Co(30)]$_{12}$ | 0.26 | 0.76 | 0.18 | 0.14 | 0.16 | - |
| [Ru(6)/Cr(3)/Co(40)]$_{12}$ | 0.14 | 0.80 | 0.13 | - | 0.15 | - |
| [Ru(6)/Cr(3)/Co(50)]$_{12}$ | 0.13 | 0.89 | 0.12 | 0.34 | 0.12 | - |

## 4. Micromagnetic Simulations

The micromagnetic simulations have been done using the mumax3 package [27] for a model system of 12 Co layers with thickness $t_{Co}$=25Å, with antiferromagnetic interlayer coupling.

Based on the magnetic measurements the saturation magnetization was set to $M_S$=1160 kA/m and the uniaxial anisotropy to $K_{mc}$= 460 kJ/m³ which represents a typical value for our samples. It is larger than the 290 kJ/m³ of the single Co layer due to the interfacial contributions. Note that these values are lower than the $\mu_0 M_S^2 = 1690$ kJ/$m^3$ required for perpendicular anisotropy. Thus, the magnetization is easy-plane. The bulk intralayer exchange stiffness was set to $A_{ex}$=17 pJ/m [25]. The interfacial RKKY exchange was set to $J_{AF} = -0.5$ mJ/$m^2$. The cell size was set to 2.5 nm which is smaller than the characteristic exchange length scale $L_{ex} = \sqrt{2A_{ex}/\mu_0 M_S^2} = 4.5$ nm. The lateral simulation cell size was set to 320nm×320nm. An array of 10x10 extra in-plane images was set, as pseudo-periodic boundary conditions, to better account for the demagnetizing field of thin film geometry.

In Fig.4 the total magnetization as a function of the applied field is shown for the multilayer and is compared with the curves for each of the constituent Co layers. The total magnetization increases linearly and reaches saturation at a value of 0.67 Tesla. This is exactly what is predicted by a two-macrospin model for which the saturation field should be $H_\parallel = 2H_{ex} = 4J_{AF}/(M_s t_{Co})$. The factor 2 accounts for the fact that each layer is coupled on both sides. The fact that the two outer layers are not coupled on both sides does not seem to affect the response to the multilayer as a whole: It does not even lead to a reduction of ($N$-1)/$N$=11/12 (where $N$ is the number of layers) as a simple linear scaling would suggest. To check this fact, we have also simulated $N$=2,4 and 8. For $N$=2 we get the expected value $H_\parallel = H_{ex} = 2J_{AF}/(M_s t_{Co})$. For $N$=4 we get 87% instead of 75% of $2H_{ex}$, and for $N$=8 we already get 98.5% instead of 87.5% of the $2H_{ex}$. However, the lack of coupling of the outer layers (n=1, 12 at Fig.5) has a significant impact on their saturation which proceeds much faster that the average linear dependence. This forces their adjacent layers (n=2, 11) to the opposite direction and so on, yielding the layer dependent approach to saturation sketched in Fig.4. In short, at a specific applied external field the symmetry axis of the scissor state varies along the film thickness. Of course, due to the symmetry the dependence of the *n*-th layer coincides with that of the (13-*n*)-th.

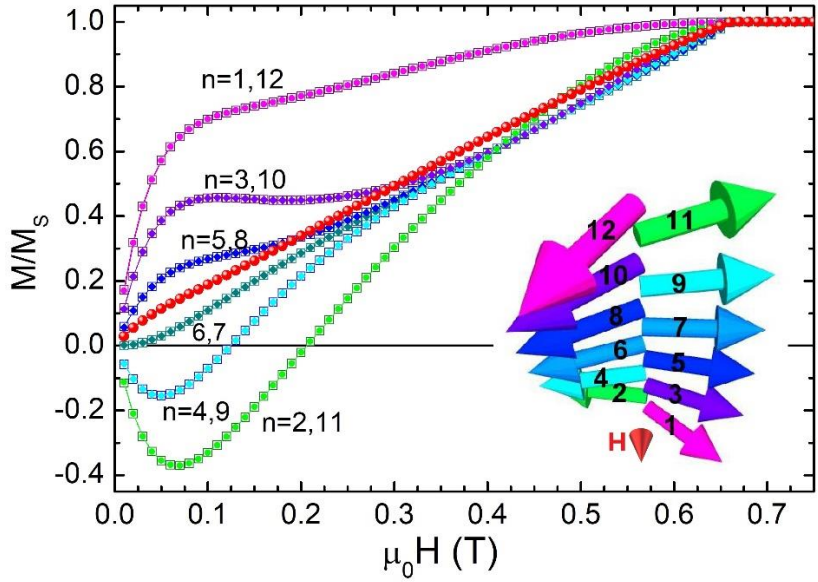

Fig.4. *Simulated magnetization curve (red circles) of a SAF consisting of 12 Co layers compared to the magnetization of each of the twelve layers. While the total magnetization is very close to the linear dependence predicted by a two-macrospin model, the individual layers have significant deviations. The inset shows the configuration for $\mu_0 H=0.23$ T. Each of the vectors is a macrospin along the direction of the moment of the layer. Coloring is done according to the cosine with the direction of the applied field (small red cone). The outer layers (1,12) that are coupled only on one side, are the first to move towards the field forcing their neighboring layers (2,11) to the opposite direction due to the AF coupling.*

The resonance frequencies are extracted following the methods and considerations described in reference [28]: for each magnetic state (different dc field value H) along the saturation curve, an exciting external field having a sinc time dependence, is applied and the resulting magnetic response is Fourier analysed. The sampling time step was set to 5 ps (frequencies up to 100 GHz) and the exciting field amplitude was $H_{rf} = 1\ mT$. The peaks of the Fourier transform correspond to the resonance modes. Setting the sinc pulse perpendicular/parallel to the dc field H (always in-plane) favours the excitation of acoustic/optical modes respectively [4]. A 2D-contour map of the Fourier transform amplitude as a function of frequency and applied field is shown in Fig.5. The plotted quantity is the amplitude of the variation of the magnetization component $\delta M$ along the direction of $H_{rf}$ divided by the $H_{rf}$ amplitude $\chi = \left( \frac{1}{H_{rf}} \cdot \frac{\delta M}{M_s} \right)$. The frequency is normalized to the value $\gamma \sqrt{2 H_{ex} H_K}$ of the zero-field optical mode. The applied field is normalized to the saturation field against the AF exchange interlayer coupling which is equal to $2H_{ex}$. On the contour maps the frequencies predicted by eqs 1 and 2 are superimposed. The logarithmic scale was chosen to magnify the contributions of weaker resonances, i.e. extra resonances apart from those predicted from eqs 1 and 2. Fig.5 shows that at low applied fields several modes of mixed character with frequencies between those of the acoustic and optical branches are exited. A gap appears near $H^*$ and mixed character excitations also appear. For the $[Ru_{10}/Co_x)]_{12}$ samples the values of $f_{cavity}/(\gamma\sqrt{2H_{ex}H_K})$ range between 0.48 to 0.56. In this region, there is a strong deviation of amplitude maximum from the curve predicted by the two macrospin model and an additional peak close to the line of the acoustic resonances. For the $[Ru_6/Cr_3/Co_x)]_{12}$ samples the corresponding values range between 0.85 to 2.5. As it was mentioned, for $f_{cavity}/(\gamma\sqrt{2H_{ex}H_K}) > 1$ greater the optical modes cannot be observed. For the samples for which the values are close to unity no significant difference between the micromagnetic and the two spin model predictions is expected as this region is far from the hybridized region.

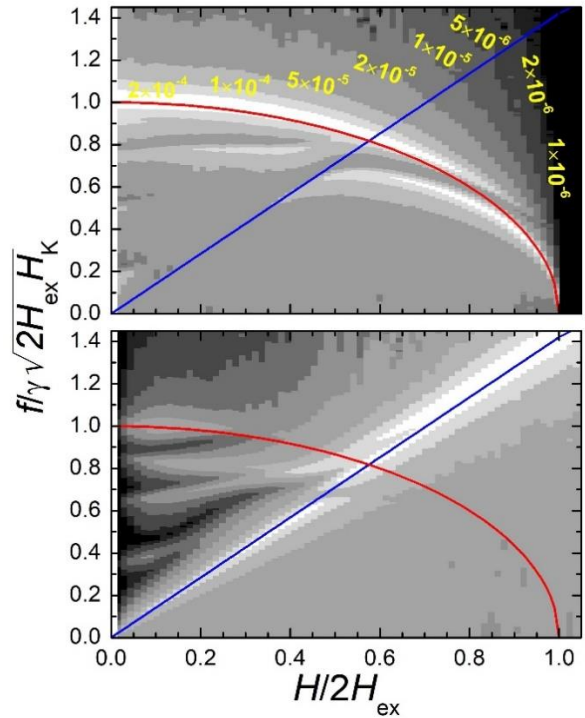

Fig.5. *Amplitude FFT transform of the magnetization component along the applied dc field as a function of frequency and dc field for a [Ru/Co)]₁₂ SAF multilayer. The plotted quantity is $\chi = \left(\frac{1}{H_{rf}} \cdot \frac{\delta M}{M_S}\right)$. The color map (between black and white) is not linear but corresponds to the $\chi = 10^{-6}, 2 \cdot 10^{-6}, 5 \cdot 10^{-6}, 10^{-5}, 2 \cdot 10^{-5}, 5 \cdot 10^{-5}, 10^{-4}, 2 \cdot 10^{-4}$. The frequency is normalized to the value $\gamma\sqrt{2H_{ex}H_K}$ of the zero-field optical mode. The applied field is normalized to the saturation field $2H_{ex}$ against the interlayer AF exchange. The frequencies predicted by the macrospin model (eqs 1 and 2) are superimposed. Red curve is used for the optical mode and blue line for the acoustic.*

Since the preferential excitation of either acoustic or optical modes, depends on the orientation of the exciting field with respect to the magnetization direction, the variation of the magnetization profile along the multilayer thickness favours the emergence of mixed modes. If we simulate a simple SAF bilayer, which does not allow for any variation of the magnetization along its thickness the mixed modes cease to appear (Fig.6).

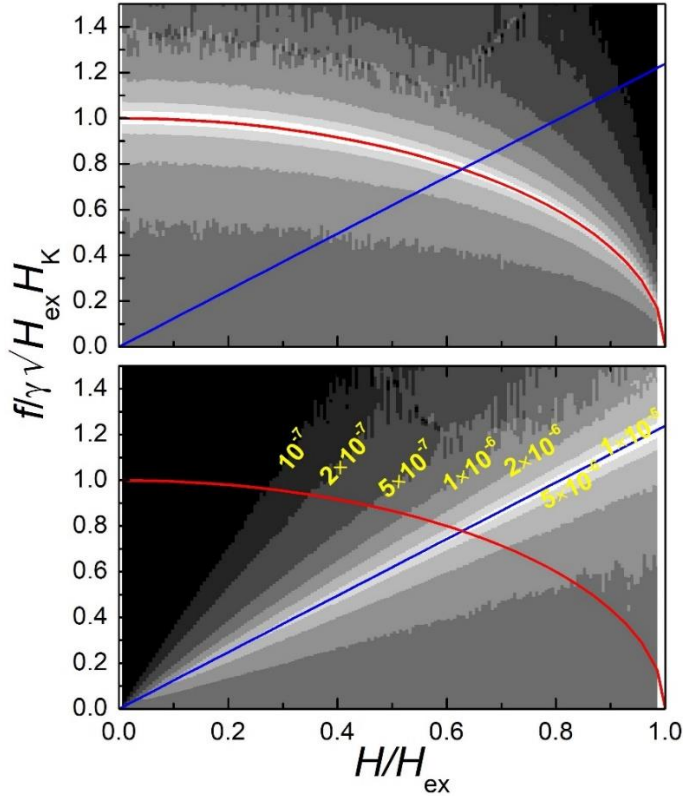

Fig.6. *Amplitude FFT transform of the magnetization component along the applied dc field as a function of frequency and dc field for a [Ru/Co]₂ SAF bilayer. The plotted quantity is* $\chi = \left(\frac{1}{H_{rf}} \cdot \frac{\delta M}{M_s}\right)$. *The color map (between black and white) is not linear but corresponds to* $\chi = 10^{-7}, 2 \cdot 10^{-7}, 5 \cdot 10^{-7}, 10^{-6}, 2 \cdot 10^{-6}, 5 \cdot 10^{-6}, 10^{-5}$. *The frequency is normalized to the value* $\gamma \sqrt{H_{ex}H_K}$ *of the zero-field optical mode. The applied field is normalized to the saturation field* $H_{ex}$ *against the interlayer AF exchange. The frequencies predicted by the macrospin model (eqs 1 and 2, but with* $2H_{ex} \rightarrow H_{ex}$*) are superimposed. Red curve is used for the optical mode and blue line for the acoustic.*

## 5. Conclusions

We have studied Ferromagnetic Resonance (FMR) in two series of $[Ru_{10}/Co_x]_{12}$ and $[Ru_6/Cr_3/Co_x]_{12}$ x=16-60Å, SAF-multilayers prepared by sputter deposition having easy-plane anisotropy and zero anisotropy within the plane.

The resonance fields of the acoustic modes are in good agreement with the values predicted by macrospin models when using the interlayer exchange and anisotropy fields independently derived by the quasistatic (VSM) magnetic measurements. The optical modes are more interesting as they can give high frequencies at zero-applied field. However, to observe optical modes by cavity FMR, $\gamma\sqrt{2H_{ex}H_K}$ must exceed the resonance frequency of the cavity. For the $[Ru_6/Cr_3/Co_x]_{12}$ series the values and of anisotropy and exchange fields were low and the optical modes were not accessible. For the $[Ru_{10}/Co_x]_{12}$ series the resonance fields of the optical modes are lower than the expected ones. We attribute this to the existence of hybridized mixed modes as the resonances for these samples appear within the region where acoustic and optical modes are expected to hybridize. The existence of inhomogeneous modes that can be described as coupled acoustic and optical modes has been previously reported [12,22] and gained renewed interest lately [4,5,21]. The coupling mechanism is related to asymmetry due to obliquely applied external magnetic field or of the sample itself. Since the preferential excitation of either acoustic or optical modes, depends on the orientation of the exitingexciting field with respect to the magnetization direction, the variation of the magnetization profile along the multilayer thickness, revealed by our micromagnetic simulations, implies the emergence of such hybridized modes.

### Acknowledgements

P. Ntetsika acknowledges funding in the framework of the programme "PhD student scholarships 2021" of the Research Committee of the University of Ioannina.